\documentclass[useAMS]{mn2e}

\usepackage{graphicx,color,soul}
\usepackage{latexsym}
\usepackage{multirow}
\usepackage{amsmath,amssymb}        
\usepackage[draft=false]{hyperref}

\title[Flip-flop behavior]{Is the flip-flop behaviour of accretion shock cones on to black holes an effect of coordinates?}

\author[A. Cruz-Osorio, F. D. Lora-Clavijo, F. S. Guzm\'an]{A. Cruz-Osorio,F. D. Lora-Clavijo,F. S. Guzm\'an \thanks{E-mail:
alejandro@ifm.umich.mx (ACO);fadulora@ifm.umich.mx (FDLC); guzman@ifm.umich.mx (FSG)} \\
	     Instituto de F\'{\i}sica y Matem\'{a}ticas, Universidad
              Michoacana de San Nicol\'as de Hidalgo.\\ Edificio C3, Cd.
              Universitaria, 58040 Morelia, Michoac\'{a}n,
              M\'{e}xico.}

\begin{document}


\date{\today}

\pagerange{\pageref{firstpage}--\pageref{lastpage}} \pubyear{2011}

\maketitle

\label{firstpage}


\begin{abstract}
We study numerically the relativistic Bondi-Hoyle accretion of an ideal gas onto a Kerr fixed background space-time on the equatorial plane with s-lab symmetry. We use both Kerr-Schild  (KS) and Boyer-Lindquist (BL) coordinates. We particularly focus on the study of the flip-flop motion of the shock cone formed when the gas is injected at supersonic speed.  The development of the flip-flop instability of the shock cone in the relativistic regime was reported recently for the first time. We reproduce the flip-flop behaviour found in the past when BL coordinates are used, and perform similar numerical experiments using horizon penetrating KS coordinates. We find that when using KS coordinates the shock cone oscillates, however such oscillations are not of the flip-flop type and their amplitude  decrease with resolution.
\end{abstract}




\section{Introduction}

Bondi-Hoyle accretion is the evolution of a homogeneously distributed gas moving uniformly toward a central compact object \cite{Hoyle}. Depending of the relation existing between the velocity of the gas and the sound speed  a shock cone can be formed. Specifically, when the fluid is supersonic, a shock cone appears, otherwise it does not.

When the compact object is rotating, an interesting phenomenon, called flip-flop instability can appear \cite{matsuda}. This effect consists in the oscillation of the shock cone orientation.  Various studies of this effect have been carried out under different conditions and for different purposes. For example, in Newtonian gravity, considering the density and velocity field to be non-uniform, they found that the shock cone might achieve a disk-like topology \cite{fryxell,taam}; also the dependence of the appearance of the flip-flop behaviour in terms of the size of the object  has been studied \cite{sawada,matsuda2,livio,taam2}.

The astrophysical relevance of the flip-flop of the shock cone is associated for example with the fluctuations observed in the X-ray emission of pulsing systems such as Vela-x1 and EXO 2030+375, that are not associated to the orbital period of the binary system \cite{taam}.

The flip-flop effect can also  happen within a relativistic  scenario. Particularly interesting is the case when the compact object is a  black hole.  This case was first reported in \cite{Donmez} and we consider this is a very important subject to be studied in detail, because the oscillations  (of  different nature) of a gas around a black hole may potentially explain the oscillatory behaviour of high energy sources, like quasi periodical oscillating sources (QPOs) or bursts when the shock cone is unstable. Moreover, we consider the treatment involving the curved background space-time to be a very important step, which uses a truly black hole space-time background instead of the commonly used pseudo-Newtonian Paczy\'nski-Wiita type potentials.

We are in the position of analyzing the case where the gas is truly allowed to enter the black hole's event horizon, this is what we do in this paper. In the relativistic Bondi-Hoyle accretion, the flip-flop instability in the equatorial plane was briefly reported by \cite{Donmez} . However, the coordinates used to describe the rotating Kerr black hole are not appropriate, because the Boyer-Lindquist coordinates are singular at the event horizon, and as a consequence it is necessary -as in the Newtonian cases- to apply  boundary conditions at an inner time-like artificial boundary outside and near the horizon. 

In this work we study the Bondi-Hoyle accretion process using the Kerr-Schild penetrating coordinates in the equatorial plane, in these coordinates the fluid naturally falls into the black hole. We also study the case of using the singular Boyer-Lindquist coordinates in order to reproduce the flip-flop effect and measure its amplitude and phase. We reproduce the morphological behaviour in  \cite{Donmez}, and measure the amplitude of the flip-flop oscillations  using  Boyer-Lindquist coordinates. However, when using KS coordinates, there are some oscillations with a frequency that is not comparable to that observed when using BL coordinates associated to the flip-flop effect; we observe that these oscillations are due to numerical artifacts because their amplitude  converges to zero when resolution is increased .

In fact, there is a precedent within the Newtonian regime where it has been found that shock cone instabilities can be associated with the implementation of boundary conditions at the surface of the accretor, and may also influence the attached or detached nature of the shock to the compact object or numerical parameters in the calculations \cite{Flogizzo}.

This paper is organized as follows. In Sec. \ref{sec:equations} we describe the system of the relativistic hydrodynamics equations and the metric functions for each coordinate system, in Sec. \ref{sec:numerical_methods}  we describe the numerical methods used for evolving  the equations, initial wind configurations and boundary conditions. In Sec. \ref{sec:results} we have a comparison of the flip-flop behaviour between Boyer-Lindquist and Kerr-Schild coordinates system and in Sec. \ref{sec:conclusions} we present a discussion of our results and some conclusions.

\section{Relativistic Hydrodynamic Equations}
\label{sec:equations}

We model the wind using the relativistic Euler equations for a perfect fluid described by the stress-energy tensor

\begin{equation}
T^{\mu \nu} =\rho_0 h u^{\mu}u^{\nu} + pg^{\mu\nu},
\end{equation}

\noindent where $\rho_0$ is the rest mass density of a fluid element, $u^{\mu}$ the 4-velocity of the fluid elements, $h=1+\epsilon+p/\rho_0$ is the specific internal enthalpy, with $\epsilon$ the rest frame specific internal energy and $p$ the pressure. The space-time background is described by the metric $g_{\mu\nu}$, and we use units where $G=c=1$ from now on. Additionally, we assume the fluid obeys an ideal gas equation of state given by

\begin{equation}
p =\rho_0 \epsilon (\Gamma-1), \label{eq:EOS}
\end{equation}

\noindent where $\Gamma$ is the adiabatic index, or the ratio between specific heats. Relativistic Euler's equations are derived from the local conservation of the rest mass density and the local conservation of the stress-energy tensor

\begin{eqnarray}
\nabla_{\mu}(\rho_0 u^{\mu}) &=& 0,\nonumber\\
\nabla_{\mu}(T^{\mu\nu}) &=& 0,\nonumber
\end{eqnarray}

\noindent where $\nabla_{\mu}$ is the covariant derivative consistent with $g_{\mu\nu}$  \cite{Misner}. It is convenient to cast these equations in a flux balance law fashion, as described in \cite{Marti,Banyouls,Font}. The only ingredient left is the space-time background written within the 3+1 decomposition of the space-time formalism. Such metric is written in general as 

\begin{equation}
ds^{2}= -(\alpha^{2} - \beta_{i}\beta^{i})dt^{2} + 2\beta_{i}dx^{i}dt + \gamma_{ij}dx^{i}dx^{j}, 
\end{equation}

\noindent where $\beta^i$ is the shift vector, $\alpha$ the lapse function and $\gamma_{ij}$ are the components of the spatial 3-metric.

The background space-time corresponds to a rotating black hole, however, since the main aim of the paper is the study of the influence of a time-like internal artificial boundary used in previous studies we consider the case of the black hole being described with BL coordinates which is the usual approach, and the case of using KS coordinates, which have the property of penetrating the black hole event horizon. In both cases we describe the space-time using spherical coordinates for the space  $x^{\mu}=(t,r,\theta,\phi)$.

In the first case, for BL coordinates the metric functions are

\begin{eqnarray}
\alpha &=&  \left( \frac{\varrho^2 \Delta}{\Sigma}  \right)^{1/2},\nonumber\\
\beta^i &=& \left(  0, 0, -\frac{2Mar}{\Sigma} \right),\nonumber\\
\gamma_{ij} &=&
\left( \begin{array}{ccc}
\frac{\varrho^2}{\Delta}&0&0 \\
0&\varrho^2&0\\
0&0&\frac{\Sigma}{\varrho^2}\sin^2\theta 
         \end{array}
\right),\nonumber
\end{eqnarray}

\noindent where

\begin{eqnarray}
\Delta &\equiv& r^2 - 2Mr + a^2,\\
\varrho^2 &\equiv& r^2 + a^2\cos^2\theta, \\
\Sigma &\equiv& (r^2+a^2)^2 - a^2\Delta\sin^2\theta,
\end{eqnarray}

\noindent and $M,a$ are the mass and the angular momentum  of the black hole respectively. 

On the other hand, in KS coordinates the metric functions are

\begin{eqnarray}
&&\alpha = \left( 1+\frac{2Mr}{\varrho^2}\right)^{-1/2},\nonumber\\
&&\beta^i = \left(  \frac{2Mr}{\varrho^2}\left( 1+\frac{2Mr}{\varrho^2}\right)^{-1}, 0, 0 \right),\nonumber\\
&&\nonumber\\
&&\gamma_{ij} =\nonumber\\
&&\left( 
	\begin{array}{ccc}
1+\frac{2Mr}{\varrho^2}&0&-a\left(1+\frac{2Mr}{\varrho^2}\right)\sin^2\theta \\
0&\varrho^2&0\\
-a\left(1+\frac{2Mr}{\varrho^2} \right)\sin^2\theta&0&\sin^2\theta [ \varrho^2+a^{2}\left(1+\frac{2Mr}{\varrho^2}\right)\sin^2\theta ]
         \end{array}
\right).\nonumber\\\nonumber
\end{eqnarray}

At this point we assume the system is cylindrically symmetric, that is, we only study the evolution of the gas on the equatorial plane and set $\theta=\pi/2$, which does not mean -as usually assumed- this is a sort of an infinitesimally thin morphology approximation.

With the space-time background metric known, relativistic Euler equations on a curved fixed background space-time in our coordinates and for the two cases considered read \cite{Marti,Banyouls,Font}

\begin{eqnarray}
\partial _{t} \vec{U} + \partial _{r} \vec{F}^{r} +\partial _{\phi} \vec{F}^{\phi} = \vec{S} - \frac{\partial_{r} \sqrt{\gamma}}{\sqrt{\gamma}} \vec{F}^{r}, \label{eq:flux_conservative}
\end{eqnarray}

\noindent where $\gamma=det(\gamma_{ij})$ is the determinant of the spatial metric, $\vec{U}$ is the vector whose entries are conservative variables that depend on the original primitive variables $\rho_0,v^{i},p,\epsilon$. The vector $\vec{F}^{i}$ contains the fluxes along the spatial directions  and $\vec{S}$ is the source vector. All these ingredients are specifically

\begin{eqnarray}
\vec{U} &=& \left[ \begin{array}{l}
D \\ 
S_{r} \\
S_{\phi} \\
\tau   
\end{array} \right]=
 \left[ \begin{array}{l}
\rho_0 W \\ 
\rho_0 h W^{2}v_{r} \\
\rho_0 h W^{2}v_{\phi} \\
\rho_0 h W^{2}  -p-\rho_0 W
\end{array} \right],\label{eq:cvars}
\\
\vec{F}^{r} &=& 
 \left[ \begin{array}{l}
\alpha D( v^{r} - \frac{\beta^{r} }{\alpha}) \\ 
 \alpha S_{r}\Big(v^{r}-\frac{\beta^{r}}{\alpha}\Big)  + \alpha p  \\
 \alpha S_{\phi}\Big(v^{r}-\frac{\beta^{r}}{\alpha}\Big)   \\
\alpha \bigg\{ \tau\Big(v^{r}-\frac{\beta^{r}}{\alpha}\Big) + pv^{r} \bigg\} 
\end{array} \right],\label{eq:fluxes}
\\
\vec{F}^{\phi} &=& 
 \left[ \begin{array}{l}
\alpha D( v^{\phi} - \frac{\beta^{\phi} }{\alpha}) \\ 
 \alpha S_{r}\Big(v^{\phi}-\frac{\beta^{\phi}}{\alpha}\Big)  \\
 \alpha S_{\phi}\Big(v^{\phi}-\frac{\beta^{\phi}}{\alpha}\Big)  + \alpha p\\
\alpha \bigg\{ \tau\Big(v^{\phi}-\frac{\beta^{\phi}}{\alpha}\Big) + pv^{\phi} \bigg\} 
\end{array} \right],\label{eq:fluxes2}
\end{eqnarray}

\begin{eqnarray}
\vec{S} = 
 \left[ \begin{array}{l}
0\\ 
 \alpha T^{\mu \nu} g_{\nu \sigma } \Gamma ^{\sigma}_{\mu r} \\
 \alpha T^{\mu \nu} g_{\nu \sigma } \Gamma ^{\sigma}_{\mu \phi} \\
\alpha  \lbrace T^{\mu t} \partial_{\mu}\alpha  - T^{\mu \nu} \Gamma^{t}_{\mu\nu} \alpha   \rbrace
\end{array} \right],\label{eq:sources}
\end{eqnarray}

\noindent where $v^i$ are the components of the velocity of the gas measured by an Eulerian observer, which are related to the spatial components of the 4-velocity of the fluid elements by $v^{i}=u^{i}/W + \beta^{i}/\alpha$, $W$ is the Lorentz factor $W=1/\sqrt{1-\gamma_{ij}v^{i}v^{j} }$; $\Gamma^{\sigma}_{\alpha\beta} $ are the Christoffel symbols obtained from the space-time metric.

Finally, the system of equations (\ref{eq:flux_conservative},\ref{eq:cvars},\ref{eq:fluxes},\ref{eq:fluxes2},\ref{eq:sources}) is a set of four equations for the variables 
$\rho_0,v^r,v^{\phi},p, \epsilon$. Thus, in order to close this system of equations, we consider the fluid obeys the equation of state (\ref{eq:EOS}).

 \section{Numerical methods}
 \label{sec:numerical_methods}

We evolve numerically the system of equations (\ref{eq:flux_conservative},\ref{eq:cvars},\ref{eq:fluxes},\ref{eq:fluxes2},\ref{eq:sources}) as a 2D initial value problem using a finite volume approximation on the equatorial plane. Specifically we use the well known high resolution shock capturing methods, with the HLLE approximate Riemann solver  \cite{hlle,hlle1}, and the minmod linear piecewise reconstructor. For the update in time we use the method of lines with a second order Runge-Kutta TVD integrator along the time direction.

\subsection{Initial data}

As initial data, we consider a homogeneous wind, that uniformly fills  the whole domain, moving on the equatorial plane along the $x$ direction with constant density and pressure. The initial velocity field $v^i$ can be expressed in terms of the asymptotic velocity $v_{\infty}$, as follows 

\begin{eqnarray}
v^{r}&=&F_{1}v_{\infty} \cos \phi+F_{2}v_{\infty} \sin \phi, \\
v^{\phi}&=&-F_{3}v_{\infty} \sin \phi +F_{4}v_{\infty} \cos \phi,
\end{eqnarray}

\noindent where

\begin{eqnarray}
F_{1}&=&\frac{1}{\sqrt{\gamma_{rr}}},\\
F_{2}&=&\frac{ F_{3}F_{4}\gamma_{\phi\phi}+F_{1}F_{3}\gamma_{r\phi}}{F_{1}\gamma_{rr}+F_{4}\gamma_{r\phi} },\\
F_{3}&=&\frac{F_{1} \gamma_{rr}+F_{4}\gamma_{r\phi}}{\sqrt{(\gamma_{rr}\gamma_{\phi\phi}-\gamma_{r\phi}^{2})( F_{1}^{2}\gamma_{rr} +F_{4}^{2}\gamma_{\phi\phi} + 2F_{1} F_{4}\gamma_{r\phi}    )} },\\
F_{4}&=&-\frac{2\gamma_{r\phi}}{\sqrt{\gamma_{rr}}\gamma_{\phi\phi}},
\end{eqnarray}

\noindent and the relation $v^2=v_iv^i=v^2_{\infty}$ is satisfied. Specifically, in the case of the BL coordinates the above expressions for the velocity field reduce to

\begin{eqnarray}
v^{r}&=&\sqrt{\gamma^{rr}}v_{\infty} \cos \phi ,\\
v^{\phi}&=&-\sqrt{\gamma^{\phi\phi}}v_{\infty} \sin \phi.
\end{eqnarray}
 
\noindent In order to choose the initial pressure profile, we introduce the asymptotical speed of sound  $c_{s\infty}$.  Once we fix the value of $c_{s\infty}$ and assume the density to be a constant $\rho_0=\rho_{ini}$, the pressure can be found from the following expression

\begin{equation}
p_{ini}=\frac{c_{s\infty}^{2}  \rho_{ini}(\Gamma-1)}{\Gamma(\Gamma-1)-c_{s \infty}^{2} \Gamma}.
\end{equation}

\noindent As we can see, in order to avoid negative and zero pressures, the condition $c_{s\infty}<\sqrt{\Gamma-1}$ has to be satisfied. Finally, the internal specific energy is reconstructed from the equation of state (\ref{eq:EOS}). 

We find useful to introduce the asymptotic Mach number in terms of $v_\infty$ and $c_{s\infty}$ as ${\cal M}_\infty=\frac{v_\infty}{c_{s\infty}}$.  This quantity determines whether the flow is supersonic or subsonic. When ${\cal M}_{\infty}$ is bigger than $1$ the flow is supersonic, otherwise the flow is subsonic.  In this paper, we are only interested in the supersonic case as we want to study the properties of the shock cone formed, in particular the possibility of flip-flop oscillations.

Other parameters we have to take into account are the black hole mass $M$ and its angular momentum $a$. In the case of the black hole mass we chose units in which $M=1$. On the other hand, we chose various representative cases for the rotation parameter $a$. In Table 1, we summarize the set of initial parameters we consider in our analysis.

\begin{table}\label{Table}
\centering
 \begin{tabular}{lcrrrrrr}    \hline
    Model & ${\cal M}_\infty$ & $\Gamma$ & $a$& $r_{EH}$ & $r_{acc}$  & $r_{exc}$ & $r_{max}$\\
    \hline
    \hline
    BL1  & 3 & 4/3 & 0.5&1.87&10&2&100\\
    BL2  & 4 & 4/3 & 0.5&1.87&5.88&2&60\\
    BL3  & 5 & 4/3 & 0.5&1.87&3.84&2&40\\
    \hline
    BL4  & 3 & 4/3 & 0.9&1.43&10&1.6&100\\
    BL5  & 4 & 4/3 & 0.9&1.43&5.88&1.6&60\\
    BL6  & 5 & 4/3 & 0.9&1.43&3.84&1.6&40\\
    \hline
    BL7  & 3 & 5/3 & 0.5&1.87&10&2&100\\
    BL8  & 4 & 5/3 & 0.5&1.87&5.88&2&60\\
    BL9  & 5 & 5/3 & 0.5&1.87&3.84&2&40\\
    \hline
    BL10  & 3 & 5/3 & 0.9&1.43&10&1.6&100\\
    BL11  & 4 & 5/3 & 0.9&1.43&5.88&1.6&60\\
    BL12  & 5 & 5/3 & 0.9&1.43&3.84&1.6&40\\
    \hline
    \hline
    KS1  & 3 & 4/3 & 0.5&1.87&10&1.2&100\\
    KS2  & 4 & 4/3 & 0.5&1.87&5.88&1.2&60\\
    KS3  & 5 & 4/3 & 0.5&1.87&3.84&1.2&40\\
    \hline
    KS4  & 3 & 4/3 & 0.9&1.43&10&0.9&100\\
    KS5  & 4 & 4/3 & 0.9&1.43&5.88&0.9&60\\
    KS6  & 5 & 4/3 & 0.9&1.43&3.84&0.9&40\\
    \hline
    KS7  & 3 & 5/3 & 0.5&1.87&10&1.2&100\\
    KS8  & 4 & 5/3 & 0.5&1.87&5.88&1.2&60\\
    KS9  & 5 & 5/3 & 0.5&1.87&3.84&1.2&40\\
    \hline
    KS10  & 3 & 5/3 & 0.9&1.43&10&0.9&100\\
    KS11  & 4 & 5/3 & 0.9&1.43&5.88&0.9&60\\
    KS12  & 5 & 5/3 & 0.9&1.43&3.84&0.9&40\\
    \hline
    \hline
  \end{tabular}
  \caption{ In this table we present the models studied in this paper where the models are labeled with BL and KS. In the initial fluid configuration we have six free parameters, in particular we choose three values of the Mach number for supersonic fluid configurations. We choose two values of the angular momentum of the black hole and two different values of the adiabatic index. Notice that the numerical excision boundary is always inside the event horizon when using KS coordinates and outside of it when using BL coordinates, and the outer boundary of the domain in the $r$ coordinate depends on the accretion radius  $r_{acc}$. The radii $r_{acc}$, $r_{exc}$ and $r_{max}$ are in units of $M$. In all the models we set the initial rest mass density to $\rho_0=10^{-6}$ and the asymptotic value of the sound speed $c_{s\infty}=0.1$.}
\end{table}

\subsection{Evolution}
The numerical domain is $\phi \in [0,2\pi)$ and $r \in [r_{exc},r_{max}]$. The election of both the interior boundary at $r=r_{exc}$ and the exterior one at $r=r_{max}$ deserves a careful analysis. First of all, the interior boundary may be space-like or time-like, depending on whether it lies inside the black hole event horizon or outside respectively. In most of previous analyses such boundary is time-like because the coordinates used are Boyer-Lindquist, which are singular at the event horizon, and as a way to avoid dealing with the coordinate singularity  there, $r_{exc}$ is chosen simply to be outside the event horizon; the price to pay with such an election is that this internal boundary requires the implementation of boundary conditions for the fluid variables in a region where the gravitational field is strong and the gradients of the variables involved are big, which eventually implies the propagation of numerical errors.

Our approach involves the use of KS horizon penetrating coordinates for a rotating black hole. What can be done is to define $r_{exc}<r_{EH}$, with $r_{EH}$ the radius of the event horizon at the equatorial plane, and defines a sphere where the light cones point toward the singularity of the black hole, which automatically implies that material particles approaching such boundary have no other fate than moving also toward the singularity and there is no need to impose a boundary condition there; this is a well known numerical technique used in numerical relativity called black hole excision \cite{SeidelSuen1992}. It is clear that even though the spatial slices penetrate the horizon we cannot include the singularity as a part of the numerical domain, and we choose $0< r_{exc}<r_{EH}$. This implies indirectly that the election of $r_{exc}$ will depend on the rotation parameter of the black hole $a$ that defines the event horizon radius $r_{EH}$.

The possibility to choose $r_{exc}$ inside the event horizon is the core of this paper, because we want to investigate whether or not the flip-flop instability of the shock cone in the relativistic Bondi-Hoyle accretion is due to the use of inappropriate coordinates.

With respect to the external boundary, the election of $r_{max}$ is also restricted. The formation of a shock cone strongly depends on the condition that on the one hand the gas has to move at supersonic speed and on the other, not all the material has to be accreted by the black hole, which can happen by simply considering $r_{acc}<r_{max}$, where $r_{acc}$ is the radius of accretion within which all the material is captured by the black hole and is defined in relativity by \cite{Petrich}

\begin{equation}
r_{acc}=\frac{M}{c_{s \infty}^{2}+v_{\infty}^{2} }.
\label{eq:aratio}
\end{equation}

\noindent The boundary conditions used at this boundary are outflow in the downstream region and we inject gas with the initial density and velocity in the upstream region \cite{matsuda,Fontaxi}.

The numerical grid is uniformly spaced in the $r$ and $\phi$ directions. In particular we use the base resolution of $\Delta r=0.033$ in $r$ direction and  $\Delta\phi=0.014$ in the $\phi$ direction for the evolution of all the models. We use a constant time step given by $\Delta t=C min(dr,d\phi)$, where $C$ is a fixed and constant Courant factor estimated empirically $C=0.25$. Finally, in order to avoid the divergences in our variables along the evolution, due to the definition of the specific enthalpy, we introduce an artificial  atmosphere, that is, we impose the rest mass density to be no smaller that the minimum value $1\times10^{-10}$.

\section{Results}
 \label{sec:results}

\subsection*{Morphology}

\begin{figure}
\includegraphics[width=8.5cm]{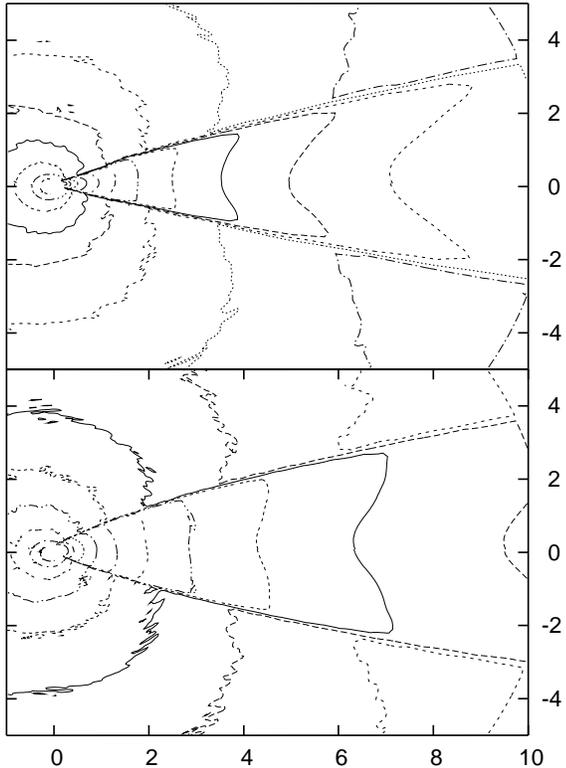}
\caption{ Here we show the contour lines of the rest mass density in order to compare the shock cone opening angle for models KS2 (top) and KS8 (bottom) in Table. We then show that the angle of the cone is bigger for bigger values of the adiabatic index (Table). The snapshot is shown for $t=10000M$. The radial coordinate in units of the accretion radius $r_{acc}$.}
\label{fig:contour}
\end{figure}

We have found similar results as done in the past, particularly, in the supersonic case, a shock cone characterized by a zone of high density shows up at the rare part of the black hole. This behaviour appears to be a sort of late-time attractor behavior. In order to compare with previous research we show various examples. As illustrative cases we show in Figure \ref{fig:contour} two examples of the shock cone morphology using KS coordinates, for two different values of the adiabatic index  $\Gamma=4/3$ and $\Gamma=5/3$. Several other cases were studied by changing the velocity of the wind and the rotation parameter of the black hole, see Table 1. The morphology of the rest mass density is consistent with previous analyses  \cite{Petrich,Fontnoaxi1,Fontnoaxi2}.

\begin{figure}
\includegraphics[width=8.5cm]{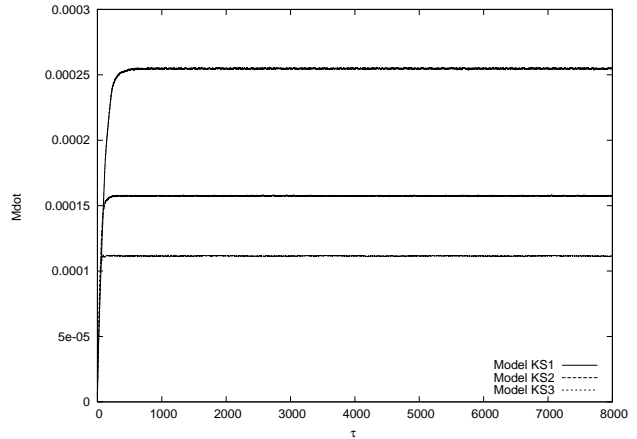}
\caption{The mass accretion rates for models KS1,KS2,KS3. We show how the accretion process depends on the velocity of the gas, being the accretion rate higher for slower winds, as expected.}
\label{fig:mdotB}
\end{figure}

Another important check is the measure of the mass accretion rate across a circle in the equatorial plane $\theta=\pi/2$, $\dot{M}=-\int_{0}^{2\pi} \alpha\sqrt{\gamma}D(v^{r}-\beta^{r}/\alpha) d\phi $, as an indicator of the accretion process during the evolution. We measure the mass accretion rate in terms of the proper time $\tau$ at a given detector. In Figure \ref{fig:mdotB} we show the accretion rate for three different values of the initial velocity of the gas in order to illustrate the dependence of the accretion rate on the velocity of the wind.  Our measurements are performed at a detector located at $r=2.1$ for three different models. The figure shows that the higher the velocity of the wind the smaller the accretion rate. Again these results are consistent with previous studies \cite{Petrich,Fontnoaxi1,Fontnoaxi2}.

\subsection*{Flip-flop behavior} 

\begin{figure}
\includegraphics[width=8.5cm]{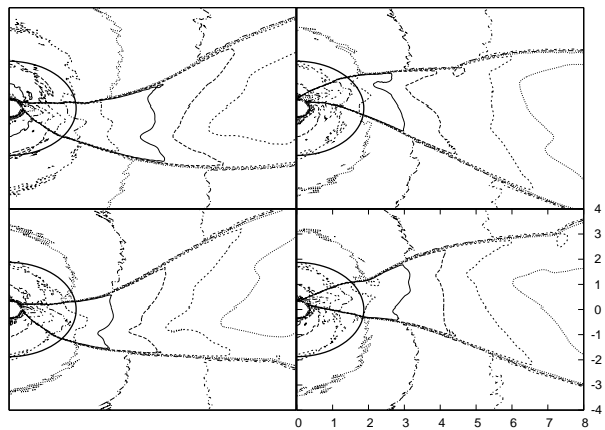}
\caption{Snapshots of the rest mass density for the model BL2 at various values of the coordinate time (top-left panel $t=7686$, top-right panel $t=8062$, bottom-left panel $t=8308$ and bottom-right panel $t=8670$) are shown. The motion of the shock cone along the angular coordinate is called the flip-flop behavior. We claim in this paper that the amplitude of such oscillations are highly dependent on the coordinates used to describe the black hole space-time, because they are singular at the event horizon. The units of the spatial coordinates are rescaled with $r_{acc}$.}
\label{fig:BL}
\end{figure}

The flip-flop motion of the shock cone consists in the oscillation of the cone itself along the angular direction. We illustrate in Fig. \ref{fig:BL} snapshots of the shock cone morphology at various times for the model BL2. It can be clearly observed that the shock cone oscillates considerably in the angular direction. Similar results in the relativistic regime were reported recently for the first time in \cite{Donmez}.

Something that called our attention is the possibility that the flip-flop behaviour might be due to the numerical implementation of boundary conditions near the event horizon, because both in  \cite{Donmez} and in Fig. \ref{fig:BL}, BL singular coordinates at the event horizon are being used. Particularly important is that in order to simulate accretion processes it is necessary to use a radial domain such that $r\in [r_{exc},r_{max}]$ with $r_{exc}>r_{EH}$, which implies that the inner boundary is time-like; moreover, if one wants to track the evolution of the gas up to regions near the black hole, the metric functions in non-penetrating coordinates tend to diverge and the gradients of the hydrodynamical variables are high. Then errors coming due to the implementation of a time-like artificial boundary near the event horizon are expected to contaminate  the simulations in the numerical domain.

\begin{figure}
\includegraphics[width=8.5cm]{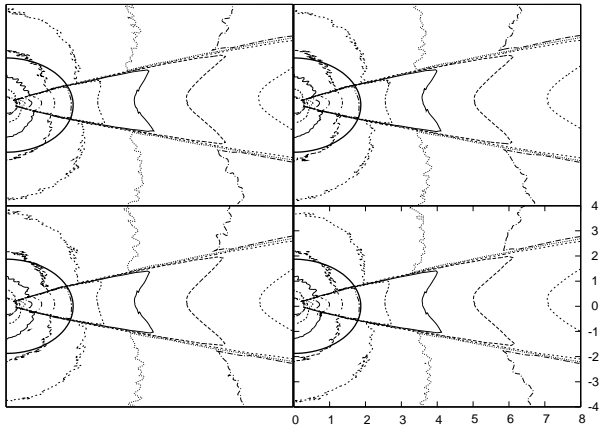}
\caption{Snapshots of the rest mass density of the shock cone, using KS coordinates for the model KS2, are shown. The corresponding times associated with each panel are: top-left panel $t=7686$, top-right panel $t=8062$, bottom-left panel $t=8308$ and bottom-right panel $t=8670$. As we can see, the motion of the shock cone is very small compared with the case in which BL coordinates are used, see Fig. \ref{fig:BL}. By using the same coordinate time values as for the BL case to expose the snapshots, we do not intend to imply that the differences in the morphology are coordinate invariant. Instead, in order to compare the motion of the shock cones in the two coordinate systems we use a detector and the proper time as measured there as explained below. The units of the spatial coordinates are rescaled with $r_{acc}$.}
  \label{fig:KS}
\end{figure}

\begin{figure}
\includegraphics[width=8.5cm]{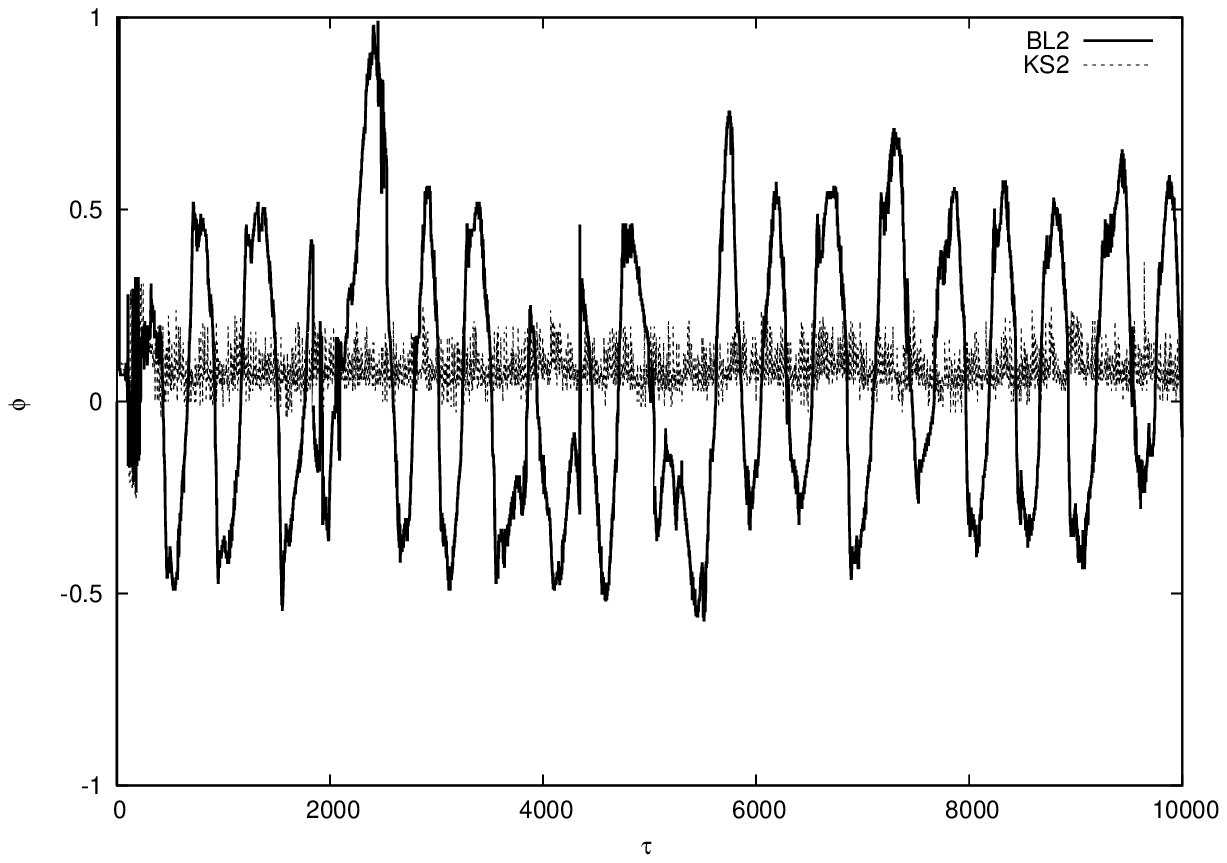}
\includegraphics[width=8.5cm]{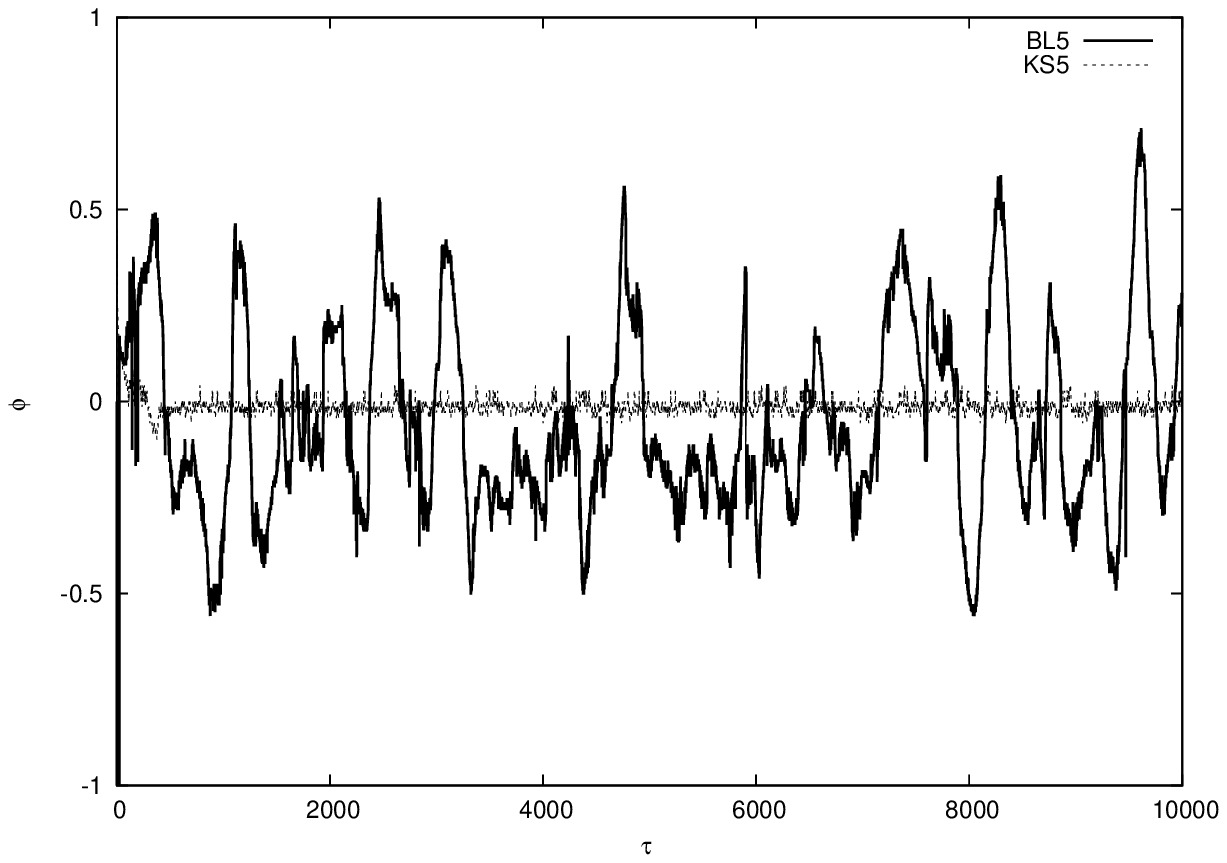}
\caption{\label{fig:angle} In these figures, we compare the  position of the maximum of the rest mass density along the angular coordinate $\phi$ during the evolution in terms of the proper time. The models we are considering to carry out this comparison are: top panel: BL2 vs KS2 and bottom panel BL5 vs KS5. In both cases the maximum of $\rho_{0} $ is measured in a detector located at $r=10M$. It is clear to see, from these figures, that when BL coordinates are considered, oscillations of high amplitude are presented, unlike in the case of KS coordinates. $\phi$ coordinate is measured in radians. }
\end{figure}

In order to investigate this, we measure the oscillations along the angular direction of the shock cone when penetrating KS coordinates are used. In Fig. \ref{fig:KS} we show snapshots of what happens for the model KS2, physically equivalent to that in Fig. \ref{fig:BL}. The oscillations are rather small compared to those observed when BL coordinates are used. In Fig. \ref{fig:angle} we show the amplitude of the oscillations when using the two coordinate systems for two representative cases in our Table. What we measure is the position of the maximum of the rest mass density along the angular coordinate in radians on a circle of radius $r=10M$; the maximum is located within the shock cone and is therefore a good quantity to be monitored; we plot it versus the proper time measured by a detector located at $r=10M$, which appears in Figs. \ref{fig:BL} and \ref{fig:KS} as a circle. 

We identify flip-flop oscillations as those that clearly show the motion of the shock cone in the angular direction. In Fig. \ref{fig:angle} we distinguish at least two important modes when using BL coordinates, one with high frequency and small amplitude and the other with big amplitude and low frequency. The later is the one associated to the flip-flop motion. The important observation here, is that when using KS coordinates, only the high frequency mode is observed, that is, the flip-flop mode does not happen. This clearly indicates that coordinates influence the motion of the shock cone, or what can be more interesting is that a true process of accretion, that is, allowing the gas to really enter the black hole, implies that the flip-flop effect does not take place.

\begin{figure}
\includegraphics[width=8.5cm]{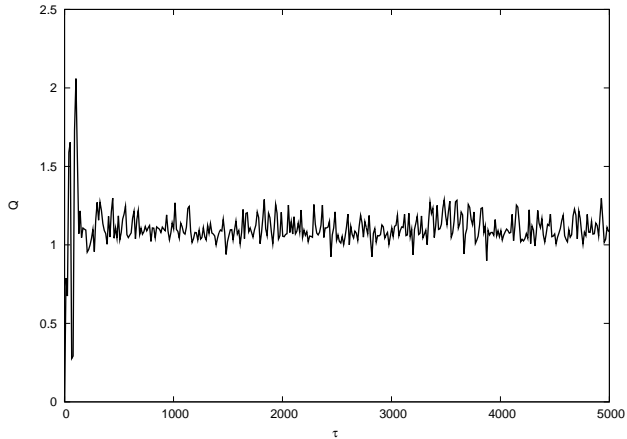}
\caption{We show the order of self-convergence $Q$ of $\rho_0$ for a fixed angle centered at the shock cone. We calculate the $L_1$ norm of the differences between the value of the density for the various resolutions. We show the self-convergence factor for resolutions  $(\Delta r_1,\Delta \phi_1 ) =(0.033,0.014) $, $(\Delta r_2,\Delta \phi_2 ) = (\Delta r_1, \Delta \phi_1)/1.5 $, and $(\Delta r_3,\Delta \phi_3 ) = (\Delta r_1, \Delta \phi_1)/(1.5)^2 $. }
\label{fig:conv}
\end{figure}

In order to check the accuracy of the numerical code, in Fig \ref{fig:conv}  we show the self-convergence of $\rho_0$ using three resolutions for the case of KS coordinates. The self-convergence was calculated for a fixed angle within the shock cone. This is a very strong test because along such line the shock cone vibrates. We have a self-convergence better than first order when the shock cone has approached a nearly stationary stage, while first order is acceptable for systems that include the evolution of shocks.

\begin{figure}
\includegraphics[width=8.5cm]{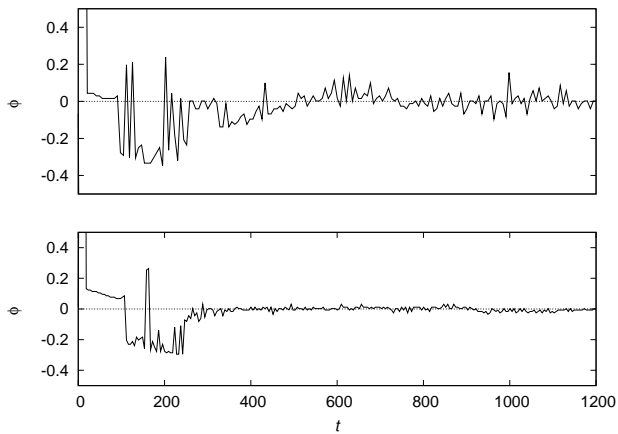}
\caption{The amplitude of the vibrations of the shock cone around the average and show how it decreases with resolution. We show only the oscillations of the maximum of the shock cone at a detector located at $r=10M$ for the two highest resolutions we used. The oscillations start at around $t\sim 500$. Before that time we see a transient lapse that corresponds to the time it takes the shock cone to approach a nearly stationary regime. The top panel shows the result using medium resolution ($\Delta r=0.022$, $\Delta \phi=0.0093$) and the bottom panel corresponds to high resolution ($\Delta r =0.0146$, $\Delta \phi =0.0062$).}
\label{fig:vibrations}
\end{figure}

In Fig. \ref{fig:angle} we show that using KS coordinates the oscillations of the shock cone are not of the flip-flop type, instead they seem to be vibrations. We go further and look at the amplitude of such oscillations. In Fig. \ref{fig:vibrations} we show the trend of the amplitude of the vibrations of the shock cone with resolution. In order to have an idea of what would happen in the continuum limit, we calculate the average of the amplitude of the vibrations for the two resolutions and measure the amplitude of the vibrations with respect to such average. The result for the exercise in Fig \ref{fig:vibrations} is that the amplitude around the average is 2.789e-2 for ($\Delta r=0.022$, $\Delta \phi=0.0093$) and 1.08e-2 for resolution ($\Delta r =0.0146$, $\Delta \phi =0.0062$).

\begin{figure}
\includegraphics[width=8.5cm]{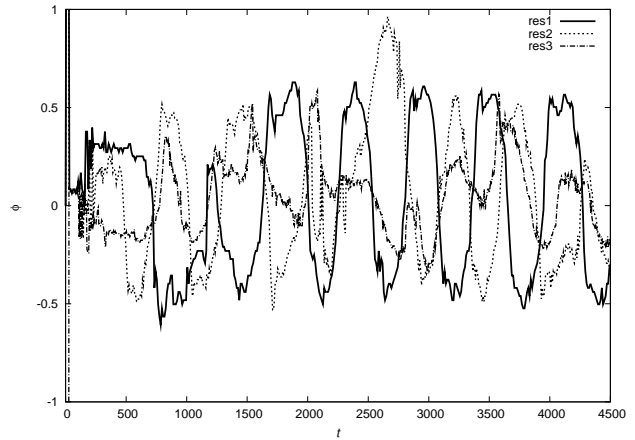}
\caption{We show the position of the maximum of the shock cone at $r=10M$ for various resolutions; we observe that the flip-flop effect persists, however the oscillations change the phase for the BL2 model.}
\label{fig:BLoscillations}
\end{figure}

In order to study the effects observed when using BL coordinates, we show in Fig. \ref{fig:BLoscillations} the position of the maximum of the density within the shock cone in time along the angular coordinate at a fixed radius. We ask whether the amplitude of the oscillations changes with resolution, and we find that the flip-flop behaviour is independent of the resolution, as one may expect it could vanish in the continuum limit.  The flip-flop behaviour is seen to be triggered with different phases for different  resolutions, which indicates that numerical discretization errors influence the appearance of the flip-flop which actually behaves as an instability.

\section{DISCUSSION AND CONCLUSIONS}
 \label{sec:conclusions}

We study numerically the relativistic Bondi-Hoyle accretion on the equatorial plane on a space-time corresponding to a rotating black hole. Our analysis is focused on the shock cone motion of a supersonic gas that is being accreted onto the rotating black hole, and its dependence on the coordinates used to describe the space-time. We compare the cases of using BL non-penetrating horizon coordinates as done in previous research and KS penetrating coordinates.

We find that when BL coordinates are used, the shock cone oscillates
in a flip-flop fashion with amplitude of the order of a radian,
whereas when using penetrating coordinates such behaviour does not take place. In fact, considering the frequency of the flip-flop oscillations when using BL coordinates, the oscillations observed when using KS coordinates do not seem to be of the same nature. Instead, in the case of KS coordinates we interpret the oscillations within the shock cone to be of  vibration or resonant type.

Furthermore, using various resolutions in our calculations, we show that the amplitude of the shock cone vibrations when using KS coordinates decreases with resolution, suggesting thus that even the vibrations are not excited in the continuum limit. Our results are naturally restricted by the s-lab symmetry that we use in our calculations, that however has been also used to identify the shock cone oscillations with -for instance- QPOs \cite{Donmez}. We suggest in this paper that perhaps such oscillations are not physically sustainable, at least within s-lab symmetry and that 3D analyses are required.

It is not possible to conclude about the condition that prevents the flip-flip effect from happening when using KS coordinates and excision inside the event horizon. We consider that allowing the matter to truly enter the black hole in the relativistic regime is the correct procedure. It may have important implications that have never been taken into account before, specially in astrophysical scenarios, where usually pseudopotentials of the Paczy\'nski-Wiita type are used to supplant a black hole, and usually it is assumed that matter will be accreted when entering the accretion radius, and that furthermore require the implementation of an artificial time-like boundary and approximate boundary conditions.


\section*{Acknowledgments}

This research is partly supported by grants: 
CIC-UMSNH-4.9 and 
CONACyT 106466.
A. C-O and F. D. L-C acknowledge support from the CONACyT.
The runs were carried out in NinaMyers computer farm.


\bsp

\label{lastpage}


\begin{thebibliography}{99}

\bibitem[Banyouls F. et al. 1997]{Banyouls}
        Banyouls F. et al., ApJ {\bf 476}, (1997) 221-231.

\bibitem[Benensohn et al. 1997]{taam2} 
	Benensohn, J. S., Lamb, D. Q.,  Taam, R. E.,1997, 
        ApJ, 478,723.	
        

\bibitem[Bondi \& Hoyle et al. 1944]{Hoyle} 
	Bondi, H., Hoyle, F., 1944, 
	Mon. Not. R. Astron. Soc. 104,273.

             
\bibitem[D\"onmez et al. 2011]{Donmez}
        D\"onmez, O., Zanotti O., Rezzolla L., MNRAS {\bf 412}, (2011) 1659-1668.    
    	

\bibitem[Einfeldt. 1988]{hlle1}
         Einfeldt, B., 1988,
        SIAM J. Numer. Anal., 25(2):294.
        
\bibitem[Foglizzo et al. 2005]{Flogizzo}
        	Foglizzo, T., Galletti, P., Ruffert, M. 2005,
	A \& A {\bf 435} 397.

\bibitem[Font \& Ib\'a\~nez et al. 1998b]{Fontnoaxi1}
        Font, J. A., Ib\'a\~nez, J. M., MNRAS {\bf 298}, (1998a) 835.         

\bibitem[Font \& Ib\'a\~nez et al. 1998a]{Fontaxi}
        Font, J. A., Ib\'a\~nez, J. M., ApJ {\bf 494}, (1998b) 297-316. 


\bibitem[Font  et al. 1999]{Fontnoaxi2}
        Font, J. A., Ib\'a\~nez, J. M., Papadopoulos, P., MNRAS {\bf 305}, (1999) 920.                  
        
\bibitem[Font et al. 2000]{Font}
        Font, J. A., Miller, M., Suen, W-M., Tobias, M., Phys. Rev.
        D {\bf 61}, (2000) 044011.      

\bibitem[Fryxell \&  Taam et al. 1988]{fryxell} 
	Fryxell, B.A., Taam, R. A., , 1988, 
        ApJ, 335,862.

\bibitem[Harten et al. 1983]{hlle}
	Harten, A., Lax, P.D. \& van Leer, B. 1083,
	SIAM Rev. 25, 35.
		
\bibitem[Livio et al. 1991]{livio} 
	Livio, M.,  Soker, N., Matsuda, T., Anzer, U., 1991, 
	Mon. Not. R. Astron. Soc. 253,633.
    
\bibitem[Marti et al. 1991]{Marti}
        Mart\'i, J. M., Iba\~nez, J. M., Miralles, J. A., Phys. Rev. D {\bf 42},
        (1991) 3794.

\bibitem[Matsuda et al. 1987]{matsuda} 
	Matsuda, T., Inoue, M., Sawada, K., 1987, 
	Mon. Not. R. Astron. Soc. 226,785.


\bibitem[Matsuda et al. 1991]{matsuda2} 
	Matsuda, T., Sekino, N., Sawada, K., Shima, E., Livio, M., Anzer, U., Borner, G., 1991, 
	A\&A 248,308
	
\bibitem[Misner et al.  1973]{Misner}
        Misner, Charles W., Thorne, Kip S., Wheeler, John A.
        Gravitation, W. H. Freeman and Company 1973.


\bibitem[Petrich et al.  1989]{Petrich}
        Petrich, L. I., Shapiro, S. L., Stark, R. F., Teulkolsky, S. A., ApJ {\bf 336}, (1989) 313.  
    
\bibitem[Sawada et al. 1989]{sawada} 
	Sawada, K., Matsuda, T., Anzer, U., Borner, G., E., Livio, 1989, 
	A\&A 231,263.

\bibitem[Seidel \& Suen 1992]{SeidelSuen1992} 
	Seidel, E. \& Suen, W-M, 1992
	Phys. Rev. Lett. {\bf 69}, 1845.
	
\bibitem[Taam \&  Fryxell et al. 1988]{taam} 
	Taam, R. A., Fryxell, B.A.,1988, 
        ApJ, 327,L73.




\end{thebibliography}
\end{document}